\documentclass[sigconf, nonacm]{acmart}

\newcommand\vldbdoi{10.14778/3554821.3554865}
\newcommand\vldbpages{3646 - 3649}
\newcommand\vldbvolume{15}
\newcommand\vldbissue{12}
\newcommand\vldbyear{2022}
\newcommand\vldbauthors{\authors}
\newcommand\vldbtitle{\shorttitle} 
\newcommand\vldbavailabilityurl{}
\newcommand\vldbpagestyle{empty} 

\newcommand{\myparagraph}[1]{\noindent \textbf{#1}}

\begin{document}
\title{Satellite Image Search in AgoraEO}

\author{Ahmet Kerem Aksoy}
\affiliation{a.aksoy@tu-berlin.de}
\affiliation{%
  \institution{TU Berlin}}

\author{Pavel Dushev}
\affiliation{pavel.dushev@sap.com}
\affiliation{%
  \institution{SAP Labs}}
\authornote{Work done while the author was at TU Berlin.}

\author{Eleni Tzirita Zacharatou}
\affiliation{elza@itu.dk}
\affiliation{%
  \institution{IT University of Copenhagen}}

\author{Holmer Hemsen}
\affiliation{holmer.hemsen@dfki.de}
\affiliation{%
  \institution{DFKI}}

\author{Marcela Charfuelan}
\affiliation{marcela.charfuelan@dfki.de}
\affiliation{%
  \institution{DFKI}}

\author{Jorge-Arnulfo Quiané-Ruiz}
\affiliation{jorge.quiane@tu-berlin.de}
\affiliation{%
  \institution{TU Berlin and DFKI}}

\author{Begüm Demir}
\affiliation{demir@tu-berlin.de}
\affiliation{%
  \institution{TU Berlin}}

\author{Volker Markl}
\affiliation{volker.markl@tu-berlin.de}
\affiliation{%
  \institution{TU Berlin and DFKI}}

\begin{abstract}
The growing operational capability of global Earth Observation (EO) creates new opportunities for data-driven approaches to understand and protect our planet. 
However, the current use of EO archives is very restricted due to the huge archive sizes and the limited exploration capabilities provided by EO platforms. 
To address this limitation, we have recently proposed MiLaN, a content-based image retrieval approach for fast similarity search in satellite image archives. 
MiLaN is a deep hashing network based on metric learning that encodes high-dimensional image features into compact binary hash codes.
We use these codes as keys in a hash table to enable real-time nearest neighbor search and highly accurate retrieval.
In this demonstration, we showcase the efficiency of MiLaN by integrating it with EarthQube, a browser and search engine within AgoraEO.
EarthQube supports interactive visual exploration and Query-by-Example over satellite image repositories.
Demo visitors will interact with EarthQube playing the role of different users that search images in a large-scale remote sensing archive by their semantic content and apply other filters.
\end{abstract}

\maketitle

\pagestyle{\vldbpagestyle}
\begingroup\small\noindent\raggedright\textbf{PVLDB Reference Format:}\\
\vldbauthors. \vldbtitle. PVLDB, \vldbvolume(\vldbissue): \vldbpages, \vldbyear.\\
\href{https://doi.org/\vldbdoi}{doi:\vldbdoi}
\endgroup
\begingroup
\renewcommand\thefootnote{}\footnote{\noindent
This work is licensed under the Creative Commons BY-NC-ND 4.0 International License. Visit \url{https://creativecommons.org/licenses/by-nc-nd/4.0/} to view a copy of this license. For any use beyond those covered by this license, obtain permission by emailing \href{mailto:info@vldb.org}{info@vldb.org}. Copyright is held by the owner/author(s). Publication rights licensed to the VLDB Endowment. \\
\raggedright Proceedings of the VLDB Endowment, Vol. \vldbvolume, No. \vldbissue\ %
ISSN 2150-8097. \\
\href{https://doi.org/\vldbdoi}{doi:\vldbdoi} \\
}\addtocounter{footnote}{-1}\endgroup


\newcommand{\eleni}[1]{\textbf{\color{blue}ETZ: #1}}
\newcommand{\begum}[1]{\textbf{\color{purple}BD: #1}}
\newcommand{\jorge}[1]{\textbf{\color{red}JQ: #1}}
\newcommand{\kerem}[1]{\textbf{\color{orange}AKA: #1}}

\section{Introduction}
\label{sec:intro}


Why do two regions in the world feel similar? What are the key characteristics that determine the fire proneness of a region? 
The stakeholders shaping the future of our planet, including scientists, practitioners, and other policy makers, typically rely on experience, precedent, and data analyzed in isolation to answer such critical questions.
In this context, the growing operational capability of global Earth Observation (EO) provides stakeholders with a wealth of information, creating tremendous opportunities for data-driven EO approaches.
However, users need a significant amount of technical expertise to validate their assumptions over large EO data archives and gain actionable insights. 

The main obstacle that non-technical users face is the limited exploration capabilities provided by EO platforms~\cite{dias}.
While they usually allow searching by geographical extent, acquisition time, or sensor type, they do not support searching by the semantic content of satellite images. This is crucial for many applications, such as mapping burnt forests or flooded residential areas. 
To help users discover relevant data in today's deluge of satellite images, image search engines that extract and exploit the image content are necessary~\cite{Sumbul2021}.  
The goal is to obtain a list of similar satellite images from a user-selected query satellite image.
For example, a user could select a portion of a satellite image acquired from a burnt forest area as a query and then let the search engine return images containing burnt forest and having similar spatial and spectral information content at a global scale.
This functionality can be crucial for climate change and ecological studies, among others.

To achieve scalable content-based image retrieval (CBIR), hashing-based approximate nearest neighbor search schemes have become a cutting-edge research topic due to their high efficiency in both storage cost and search retrieval speed~\cite{Sumbul2021}. 
Furthermore, the success of deep neural networks in image feature learning has inspired research on developing deep learning-based hashing methods (i.e. deep hashing methods). 
In our recent work, we developed several deep hashing methods for CBIR that embed high-dimensional image features into compact binary hash codes based on suitable loss functions~\cite{MiLAN}.
We use these binary codes as keys in a hash table, thereby enabling real-time nearest neighbor search.

We have also developed the BigEarthNet archive~\cite{sumbul_bigearthnet-mm_2021}, a large-scale multi-label benchmark archive for remote sensing image classification and retrieval.
We annotated each image in BigEarthNet with multi-labels that describe the different land cover types using the CORINE Land Cover (CLC) map of 2018.
Yet, this is only a part of the complete story: Users also need (i)~query tools that provide flexible label-based filtering when navigating the BigEarthNet data, and (ii)~to obtain intuitive visualizations that summarize the distribution of land cover types in a given area of interest to derive meaningful insights.

We present \emph{EarthQube}; a system that allows users to query and visualize satellite data efficiently and reverse search for satellite images based on their semantic content. 
Specifically, EarthQube supports the fast search for highly similar images given an existing satellite image.
Through a user-friendly interface, our demonstration lets users select an image of interest and obtain a ranked list of similar satellite images from a search index that contains all the images in the BigEarthNet data archive.
Furthermore, our demonstration lets users filter satellite images based on their land cover labels and provides an intuitive visualization of the occurrence of different labels in a given area.
Last but not least, our demonstration also supports standard filter operations, such as searching by geographical extent and acquisition date.

We note that while this demonstration focuses on scalable EO image search, this work is part of our larger AgoraEO vision~\cite{agoraeo}.
AgoraEO provides the technical EO infrastructure on top of Agora~\cite{agora}, a data infrastructure for AI innovation. In more detail, AgoraEO aims at supporting EO ecosystems where one can offer, discover, combine, and efficiently execute EO-related assets, such as datasets, algorithms, and tools, to get data-driven insights. 
EarthQube is a browser and search engine within AgoraEO providing efficient and easy access to BigEarthNet data.

\section{Technical Overview}
\label{sec:approarch}
We start by briefly presenting BigEarthNet, the benchmark archive that we developed and use in our demo scenarios.
Then, we present MiLaN, which is the core technology behind EarthQube.
%

\subsection{The BigEarthNet Archive}
\label{sec:bigearthnet}
The BigEarthNet archive\footnote{\url{https://bigearth.net/}} is a large-scale benchmark archive consisting of 590,326 pairs of Sentinel-1 and Sentinel-2 satellite images acquired from $10$ European countries (i.e., Austria, Belgium, Finland, Ireland, Kosovo, Lithuania, Luxembourg, Portugal, Serbia, Switzerland) between June 2017 and May 2018~\cite{sumbul_bigearthnet-mm_2021}. 
The Sentinel-2 satellite constellation acquires multispectral images with 13 spectral bands and varying spatial resolutions.
BigEarthNet excludes the 10th band because it does not embody surface information, thus keeping 12 bands per image.
Each BigEarthNet Sentinel-2 image is a section of: (i) $120\times120$ pixels for 10m bands; (ii) $60\times60$ pixels for 20m bands; and (iii) $20\times20$ pixels for 60m bands. 
The Sentinel-1 satellite constellation acquires synthetic-aperture radar data. 
BigEarthNet Sentinel-1 images contain dual-polarized information channels (VV and VH) with a spatial resolution of 10m and are based on the interferometric wide swath mode, which is the main acquisition mode over land. 
Each pair of images in BigEarthNet is annotated with multi-labels provided by the CLC map of 2018 based on its thematically most detailed Level-3 class nomenclature. 
For details about BigEarthNet, we refer the reader to ~\cite{sumbul_bigearthnet-mm_2021}.

\subsection{The MiLaN Approach}
\label{sec:cbir}

\begin{figure}[tb]
    \centering
    \includegraphics[width=\linewidth]{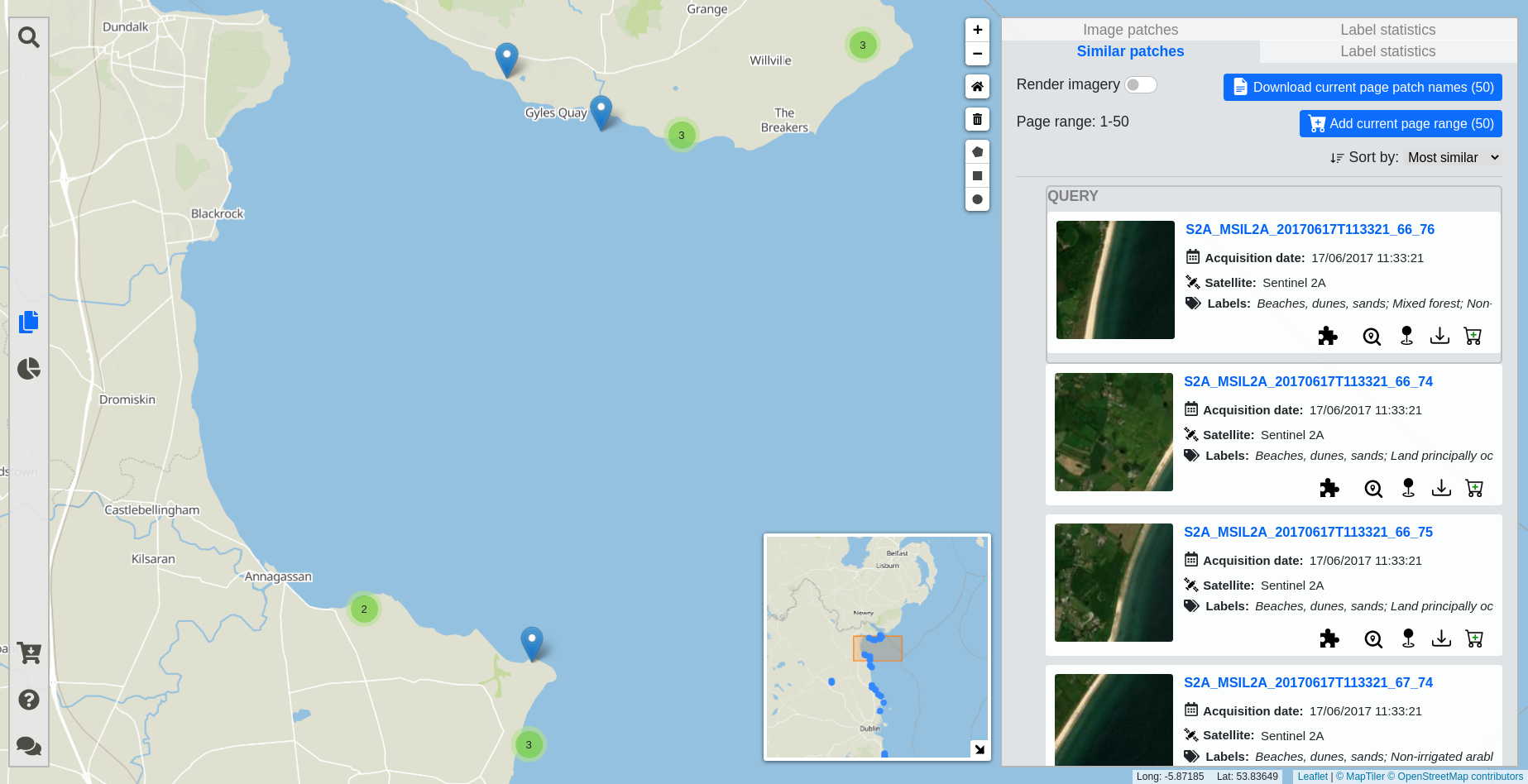}
    \caption{Content-based Image Retrieval in EarthQube.}
    \label{fig:cbir-layout}
\end{figure}

Given a query image, we aim to retrieve the most similar satellite images to the query from huge data archives in a highly time-efficient manner. 
For example, given an image that depicts a beach, we want to find images of similar beaches in different locations, as shown in Figure~\ref{fig:cbir-layout}.
To enable image indexing and scalable search, we apply deep hashing to BigEarthNet images using our recent metric learning-based deep hashing network (MiLaN)~\citep{MiLAN}.
MiLaN simultaneously learns: (i) a semantic-based metric space for effective feature representation; and (ii) compact binary hash codes for scalable search. 
To train MiLaN, we use three loss functions: (i) the triplet loss function to learn a metric space where semantically similar images are close to each other and dissimilar ones are separated; (ii) the bit balance loss function that forces the hash codes to have a balanced number of binary values (i.e., each bit has a $50\%$ chance to be activated) and makes the different bits independent from each other; and (iii) the quantization loss function that mitigates the performance degradation of the generated hash codes through binarization on the deep neural network outputs. 
As proven in~\citep{MiLAN}, the learned hash codes based on the above loss functions can efficiently characterize the complex semantics in satellite images.

After obtaining the binary hash codes of the archive images, we generate a hash table that stores all images with the same hash code in the same hash bucket. 
Then, we perform image retrieval through hash lookups, i.e., we retrieve all images in the hash buckets that are within a small hamming radius of the query image.

In this demonstration, we integrate MiLaN into EarthQube, thus allowing users to perform fast image-based similarity search on the BigEarthNet data archive.

\section{EarthQube}
\label{sec:system}
We now introduce EarthQube by first explaining its interface and then outlining its overall architecture.
Finally, we describe the integration of MiLaN to enable CBIR. 

\begin{figure*}[t]
    \centering
    \includegraphics[width=\linewidth]{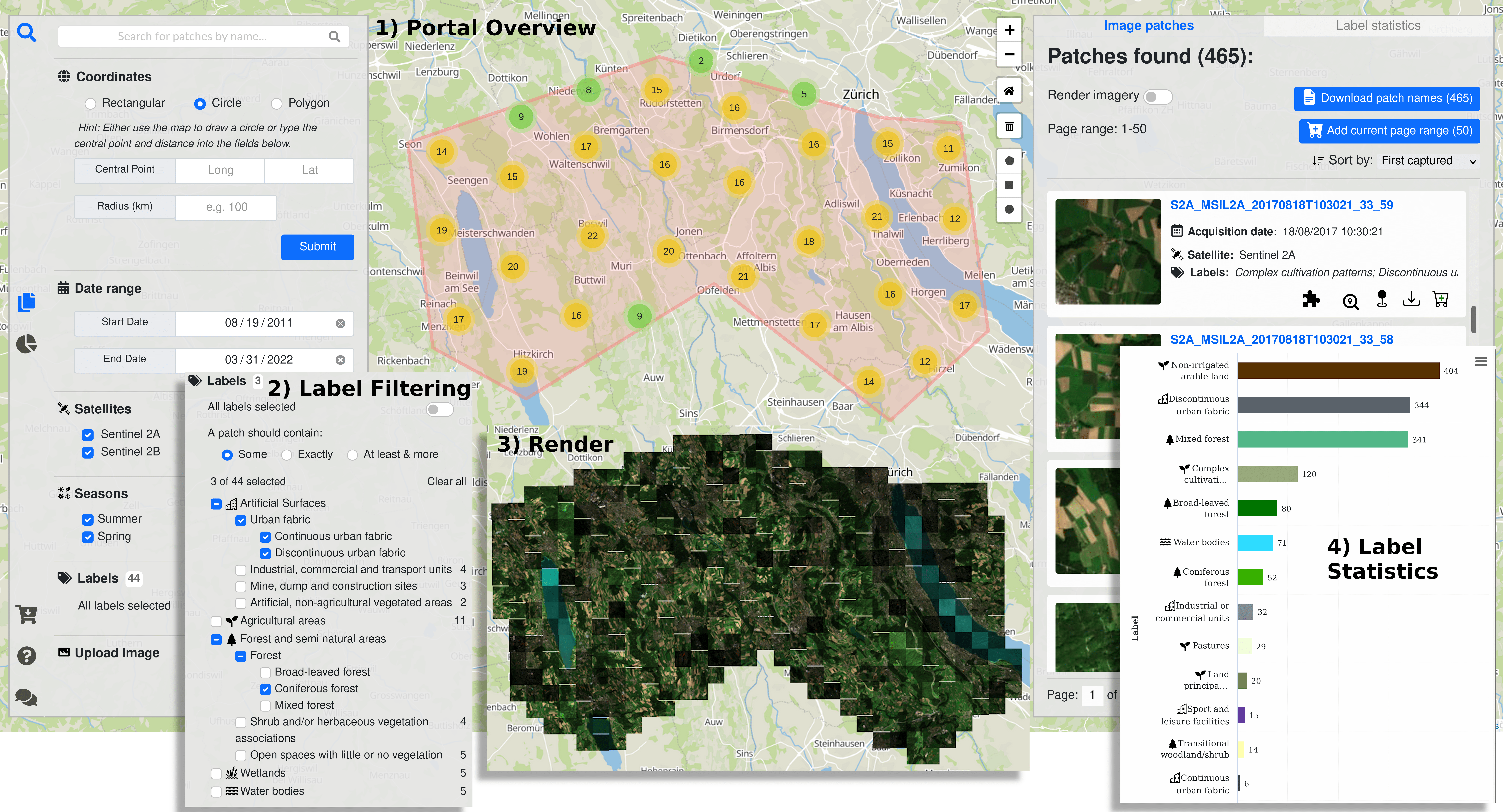}
    \caption{The visual interface of EarthQube. Users can: 1) overview the portal; 2) select query labels; 3) render result images; 4) see land cover class distribution for the query.}
    \label{fig:overall-layout-bigearthnet-portal}
\end{figure*}

\subsection{User Interface}
\label{sec:interface}
The visual interface of EarthQube is composed of a map rendering component (see Figure~\ref{fig:overall-layout-bigearthnet-portal}).
EarthQube overlays various menus and panels on the map for easy configuration and hides them when not needed for better map navigation. 
Users perform operations on the map (e.g., zoom in/out) through mouse interactions.

\vspace{0.1cm}
\myparagraph{Query Panel.} Users can issue queries through the main search menu (left side of the map in Figure~\ref{fig:overall-layout-bigearthnet-portal}-1).  
Specifically, the \emph{coordinates} subsection allows users to define a geospatial area by choosing a shape (i.e., rectangle or circle) and manually typing the area coordinates. Alternatively, users can draw an arbitrary rectangle, circle, or polygon directly on the map.    
In addition, users can filter the data based on the acquisition date range, satellites, seasons, and labels (land cover classes).

Users can control the labels using a switch button, which is initially turned on (i.e.,~no label-based filtering applies).
Turning the button off provides complete control over the label filtering criteria, as shown in Figure~\ref{fig:overall-layout-bigearthnet-portal}-2.
EarthQube groups the labels in a three-level hierarchy following the structure of the CLC land cover classes nomenclature.
Furthermore, it supports three filtering operators: \emph{Some}, \emph{Exactly}, and \emph{At least \& more}.
The \emph{Some} operator retrieves all relevant images that have at least one of the selected labels. 
For example, to retrieve images with forests, the user can select the Level-2 class \emph{Forest} that comprises of three types of Level-3 forest labels (i.e., \emph{Broad-leaved}, \emph{Coniferous}, and \emph{Mixed}).
The \emph{Exactly} operator returns images with the exact same labels as the selected ones.
This can be useful when a user is looking for very specific information, e.g. finding all airports in a provided area.
The \emph{At least \& more} operator retrieves images that have all the selected labels and potentially some additional ones.
For example, if a user is looking for sea or ocean beaches located near coniferous forests, then she is mainly interested in the labels \emph{Coniferous forest}, \emph{Beaches, dunes, sands}, and \emph{Sea and ocean}.
However, images with some additional labels, such as \emph{Bare rock} or \emph{Coastal lagoons}, could also be relevant.  
Overall, thanks to its expressive operators and the easy-to-follow hierarchical layout of the labels, EarthQube provides a powerful tool for querying EO data based on land cover classes. 

Finally, the last subsection of the query panel allows users to upload a BigEarthNet image and search for similar images in the archive using our deep hashing based index. 

\myparagraph{Map View.} 
The map displays the locations of the retrieved images as markers (zoomed-in view) and marker cluster groups (zoomed-out view). 
Markers have several features, such as hovering animations, tooltips, pop-ups, and pinpointers.
Specifically, hovering over a marker changes its color and shows its labels in a tooltip, while clicking on the marker opens a popup that contains metadata.
The pop-up also exposes a button that locates the image in the result panel that we describe next.
Furthermore, the user can choose a set of markers to pinpoint on the map.
Lastly, the bottom right of the screen shows a minimap (see Fig.~\ref{fig:cbir-layout}), which can be toggled on or off and allows users to keep an overall perspective even when they are zoomed into a particular area.
As a next step in the visual exploration, we allow users to render RGB images directly on the map, as shown in Figure~\ref{fig:overall-layout-bigearthnet-portal}-3.

\vspace{0.1cm}
\myparagraph{Result Panel.}
The result panel (right side of the map in Figure~\ref{fig:overall-layout-bigearthnet-portal}) presents metadata, additional features, and label statistics regarding the latest retrieval. 
It consists of two views: \emph{Image patches} and \emph{Label statistics}.
The top of the panel in the \emph{Image patches} view shows the total number of image patches that match the query criteria.
Furthermore, it allows users to enable image rendering on the map (up to 1000 images), download the names of the retrieved images as a plain text file, and add the current page range of images (up to 50) to the download cart.
The cart allows users to combine images from different searches and download them together as a single collection. 
The window below displays the full list of images.
Each image has a brief description and five buttons that allow to: (i) retrieve similar images, (ii) navigate to the image on the map, (iii) pinpoint the image, (iv) download the image as a zip, and (v) add the image to the download cart.

The view \emph{Label statistics} summarizes the occurrence of land cover labels in the retrieved images, which is a unique feature of EarthQube. 
Specifically, as shown in Figure~\ref{fig:overall-layout-bigearthnet-portal}-4, it consists of a bar chart that shows the number of occurrences of each label present in the retrieval.
To facilitate the identification of dominant land types in a given area, we  map each label to a predefined color that is representative of the land cover type.  

To display the results of a similarity search, EarthQube opens two new tabs for the image patches and label statistics, respectively.
These views are the same as described above, with the only difference that the image patches view displays the query image at the top in addition to the retrieved similar images (see Fig.~\ref{fig:cbir-layout}).

\balance

\subsection{System Architecture}
EarthQube follows a three-tier architecture consisting of a data tier, a back-end server, and a user interface. As we discussed the user interface in Section~\ref{sec:interface}, here we focus on the remaining two tiers. 

\myparagraph{Data Tier.} EarthQube uses MongoDB as a database server to store four data collections: (i) metadata, (ii) image data, (iii) rendered images, and (iv) user feedback.
The \emph{metadata} collection is central to EarthQube as it enables efficient search and retrieval of images based on their geospatial coordinates and other attributes. 
Specifically, metadata documents have a \emph{location} attribute that represents the bounding rectangle of an image and a \emph{properties} attribute that encompasses other queryable image features, such as the image name, labels, season, and acquisition date.  
To improve query performance, we index the \emph{location} attribute using MongoDB's built-in 2D geohashing index.
Furthermore, to improve the performance of label-based filtering, we map each (potentially multi-word) CLC label to an ASCII character, thereby avoiding the manipulation of long strings.
The \emph{image data} collection stores the actual binary representations of the 12 bands of the BigEarthNet images.
Each document has an \emph{image patch name} attribute that serves as primary key and is automatically indexed by MongoDB.
The \emph{rendered images} collection contains the binary representations of the rendered displayable images. 
We acquire those images by combining the RGB bands.
Finally, the collection \emph{feedback} stores anonymous user-provided text feedback, such as public reactions and comments. 

\myparagraph{Back-end Server.} The back-end server provides the means to submit geospatial queries, filter the images based on different search criteria, and perform CBIR.
To this end, EarthQube invokes different services that validate and process the user query. 

\subsection{Integrating with MiLaN}
To provide the CBIR functionality, we infer a 128-bit binary hash code for each image in the BigEarthNet archive using MiLaN (see Section~\ref{sec:cbir}).
EarthQube supports both querying by an existing archive image and by an external one.   
To perform a similarity search based on an archive image, we maintain an in-memory hash table that maps each image patch name to the corresponding binary code.
For queries based on an external image, the deep learning model produces a binary code for the query on-the-fly.
Given the binary code of the query image, EarthQube retrieves all images with binary codes within a small hamming radius.
Finally, the back-end server further processes the retrieved images before displaying them on the user interface.

\section{Demonstration}
\label{sec:demo}
Visitors can: (i)~interact with the BigEarthNet satellite images through our unified, easy-to-use dashboard, and (ii)~select or upload satellite images to search for similar satellite images backed by machine learning methods. They can also explore Europe and its land cover through querying and visualizing satellite images. In particular, visitors will play the following scenarios:

\myparagraph{Label-based Exploration.}
Visitors can search for industrial areas adjacent to inland water bodies using the label filtering functionality to detect possible water pollution by industrial waste in 10 different European countries. 
By inspecting the label statistics view, visitors can discover other land cover classes that fit the query description. 
They may then find out that certain areas include land principally occupied by agriculture whose irrigation may come from nearby polluted water bodies.

\myparagraph{Spatial Exploration and Query-by-Existing-Example.}
Visitors can search for urban areas or vegetation in 10 European countries that are typical of a certain region. 
For example, visitors can submit a geospatial query covering the southwestern tip of Portugal.
Then, they can visualize the images in the query area using the render functionality.
Finally, they can select an image and perform content-based image retrieval to display similar images in the 10 countries.

\myparagraph{Query-by-New-Example.}
Sentinel satellites constantly collect new images of earth's surface. Unfortunately, these newly collected images do not have any land cover class labels in the metadata. Therefore, visitors can upload such images to EarthQube to search for other images with similar semantic content. 
Based on the semantic search results, one could design an automatic labeling process. 

\begin{acks}
\small{This work is funded by the European Research Council through the ERC-2017-STG BigEarth Project (Grant 759764) and the German Ministry for Education \& Research as BIFOLD - Berlin Institute for the Foundations of Learning \& Data (ref. 01IS18025A and 01IS18037A).}
\end{acks}

\bibliographystyle{ACM-Reference-Format}
\bibliography{references}

\end{document}